\documentclass[10pt,a4paper]{article}
\usepackage{amsmath,amssymb,latexsym}
\usepackage{mathrsfs,graphicx,subfigure}

\def\AFOUR{%
\setlength{\textheight}{8.5in}%
\setlength{\textwidth}{5.75in}%
\setlength{\topmargin}{-0.375in}%
\hoffset=-.5in%
\renewcommand{\baselinestretch}{1.17}%
\setlength{\parskip}{6pt plus 2pt}%
}

\AFOUR                                           

\expandafter\ifx\csname amssym.def\endcsname\relax \else\endinput\fi
\expandafter\edef\csname amssym.def\endcsname{%
       \catcode`\noexpand\@=\the\catcode`\@\space}
\catcode`\@=11
\def\undefine#1{\let#1\undefined}
\def\newsymbol#1#2#3#4#5{\let\next@\relax
 \ifnum#2=\@ne\let\next@\msafam@\else
 \ifnum#2=\tw@\let\next@\msbfam@\fi\fi
 \mathchardef#1="#3\next@#4#5}
\def\mathhexbox@#1#2#3{\relax
 \ifmmode\mathpalette{}{\m@th\mathchar"#1#2#3}%
 \else\leavevmode\hbox{$\m@th\mathchar"#1#2#3$}\fi}
\def\hexnumber@#1{\ifcase#1 0\or 1\or 2\or 3\or 4\or 5\or 6\or 7\or 8\or
 9\or A\or B\or C\or D\or E\or F\fi}

\font\tenmsa=msam10
\font\sevenmsa=msam7
\font\fivemsa=msam5
\newfam\msafam
\textfont\msafam=\tenmsa
\scriptfont\msafam=\sevenmsa
\scriptscriptfont\msafam=\fivemsa
\edef\msafam@{\hexnumber@\msafam}
\mathchardef\dabar@"0\msafam@39
\def\dashrightarrow{\mathrel{\dabar@\dabar@\mathchar"0\msafam@4B}}
\def\dashleftarrow{\mathrel{\mathchar"0\msafam@4C\dabar@\dabar@}}

\def\ulcorner{\delimiter"4\msafam@70\msafam@70 }
\def\urcorner{\delimiter"5\msafam@71\msafam@71 }
\def\llcorner{\delimiter"4\msafam@78\msafam@78 }
\def\lrcorner{\delimiter"5\msafam@79\msafam@79 }
\def\yen{{\mathhexbox@\msafam@55}}
\def\checkmark{{\mathhexbox@\msafam@58}}
\def\circledR{{\mathhexbox@\msafam@72}}
\def\maltese{{\mathhexbox@\msafam@7A}}
\def\circledS{{\mathhexbox@\msafam@73}}

\font\tenmsb=msbm10
\font\sevenmsb=msbm7
\font\fivemsb=msbm5
\newfam\msbfam
\textfont\msbfam=\tenmsb
\scriptfont\msbfam=\sevenmsb
\scriptscriptfont\msbfam=\fivemsb
\edef\msbfam@{\hexnumber@\msbfam}
\def\Bbb#1{{\fam\msbfam\relax#1}}
\def\widehat#1{\setbox\z@\hbox{$\m@th#1$}%
 \ifdim\wd\z@>\tw@ em\mathaccent"0\msbfam@5B{#1}%
 \else\mathaccent"0362{#1}\fi}
\def\widetilde#1{\setbox\z@\hbox{$\m@th#1$}%
 \ifdim\wd\z@>\tw@ em\mathaccent"0\msbfam@5D{#1}%
 \else\mathaccent"0365{#1}\fi}
\font\teneufm=eufm10
\font\seveneufm=eufm7
\font\fiveeufm=eufm5
\newfam\eufmfam
\textfont\eufmfam=\teneufm
\scriptfont\eufmfam=\seveneufm
\scriptscriptfont\eufmfam=\fiveeufm
\def\frak#1{{\fam\eufmfam\relax#1}}

\csname amssym.def\endcsname

\makeatletter
\def\section{\@startsection {section}{1}{\z@}{-3.5ex plus -1ex minus
 -.2ex}{2.3ex plus .2ex}{\large\sc}}
\def\subsection{\@startsection{subsection}{2}{\z@}{-3.25ex plus -1ex minus
 -.2ex}{1.5ex plus .2ex}{\normalsize\sc}}
\makeatother

\makeatletter
\@addtoreset{equation}{section}

\makeatother

\newcommand{\nc}{\newcommand}
\newcommand{\rnc}{\renewcommand}

\nc{\chap}[1]{{\clearpage}%
\begin{center}%
{\noindent\underline{\large\sc #1}}{\addcontentsline{toc}{section}{#1}}%
\end{center}%
{\vspace*{0.3cm}}}

\nc{\subs}[1]{{\vspace*{0.2cm}}%
{\noindent\underline{\small\sc
#1}}%
{\vspace*{0.2cm}}}

\nc{\be}{\begin{equation}}
\nc{\ee}{\end{equation}}
\nc{\bea}{\begin{eqnarray}}
\nc{\eea}{\end{eqnarray}}

\nc{\trac}[2]{{\textstyle\frac{#1}{#2}}}

\nc{\ex}[1]{\mbox{e}^{\,\textstyle#1}}

\nc{\CC}{\Bbb{C}}
\nc{\HH}{\Bbb{H}}
\nc{\PP}{\Bbb{P}}
\nc{\RR}{\Bbb{R}}
\nc{\ZZ}{\Bbb{Z}}
\nc{\II}{\Bbb{I}}
\nc{\EE}{\Bbb{E}}
\nc{\TT}{\Bbb{T}}
\nc{\DD}{\mathrm{I}\!\mathrm{D}}

\rnc{\d}{\delta}
\nc{\eps}{\epsilon}
\nc{\om}{\omega}

\nc{\symx}{\circledS}
\newsymbol\smallsmile 1360
\newsymbol\smallfrown 1361
\nc{\ad}{\mathop{\mbox{ad}}\nolimits}
\nc{\tr}{\mathop{\mbox{tr}}\nolimits}
\nc{\Tr}{\mathop{\mbox{Tr}}\nolimits}
\nc{\Det}{\mathop{\mbox{Det}}\nolimits}
\rnc{\det}{\mathop{\mbox{det}}\nolimits}
\nc{\rk}{\mathop{\mbox{rk}}\nolimits}
\nc{\del}{\partial}
\nc{\diag}{\mathop{\mbox{diag}}\nolimits}
\nc{\ra}{\rightarrow}
\nc{\Ra}{\Rightarrow}
\nc{\LRa}{\Leftrightarrow}
\nc{\lra}{\leftrightarrow}
\nc{\ot}{\otimes}
\rnc{\ss}{\subset}
\nc{\nul}{\noindent\underline}
\nc{\non}{\nonumber\\}
\nc{\mat}[4]{\left(\begin{array}{cc}#1&#2\\#3&#4\end{array}\right)}
\rnc{\lg}{\frak{g}}
\nc{\G}[3]{\Gamma^{#1}_{\;{#2}{#3}}}
\nc{\nam}{\nabla_{\mu}}
\nc{\nan}{\nabla_{\nu}}
\nc{\dx}{\dot{x}}
\nc{\tx}{\tilde{x}}
\nc{\dtx}{\dot{\tilde{x}}}
\nc{\te}{\tilde{e}}
\nc{\dte}{\dot{\tilde{e}}}
\nc{\dxl}{\dot{x}^{\la}}
\nc{\dxm}{\dot{x}^{\mu}}
\nc{\dxn}{\dot{x}^{\nu}}
\nc{\ddx}{\ddot{x}}
\nc{\ddxm}{\ddot{x}^{\mu}}
\nc{\ddxn}{\ddot{x}^{\nu}}
\nc{\dxi}{\dot{\xi}}
\nc{\ddxi}{\ddot{\xi}}
\nc{\lsf}{\ell_s^{\mathrm{eff}}}
\nc{\lpf}{\ell_p^{\mathrm{eff}}}
\nc{\sqg}{\sqrt{g^{11}}}

\nc{\bpm}{\begin{pmatrix}}
\nc{\epm}{\end{pmatrix}}

\nc{\red}[1]{{\color{red}#1}}
\nc{\dd}{d}

\nc{\N}{\mathcal{N}}
\nc{\V}{\mathcal{V}}

\begin{document}
\vspace*{1cm}

\begin{center}
{\Large{\textsc{Horizon Shells and\\[.4cm] BMS-like Soldering Transformations}}}
\end{center}
\vspace{.5cm}
\begin{center}
\large{\textsc{Matthias Blau}${}^1$ and \large{\textsc{Martin O'Loughlin}}$^{2}$\\[.8cm]
$^{1}$Albert Einstein Center for Fundamental Physics\\
Institute for Theoretical Physics, Bern University\\
Sidlerstrasse 5, 3012 Bern, Switzerland}\\[.5cm]
$^{2}$ University of Nova Gorica\\
Vipavska 13,  5000 Nova Gorica, Slovenia
\end{center}
\vspace{1cm}

We revisit the theory of null shells in general relativity, with a
particular emphasis on null shells placed at horizons of black holes. We
study in detail the considerable freedom that is available in the
case that one solders two metrics together across null hypersurfaces
(such as Killing horizons) for which the induced metric is invariant
under translations along the null generators. In this case the group
of soldering transformations turns out to be infinite dimensional,
and these solderings create non-trivial horizon shells containing both
massless matter and impulsive gravitational wave components. 
We also rephrase this result in the language of Carrollian symmetry groups.
To illustrate this phenomenon we discuss in detail the example of shells
on the horizon of the Schwarzschild black hole (with equal
interior and exterior mass), uncovering a rich classical structure at the horizon 
and deriving an explicit expression for the
general horizon shell energy-momentum tensor.  In the special case of
BMS-like soldering supertranslations we find a conserved shell-energy
that is strikingly similar to the standard expression for asymptotic BMS supertranslation
charges, suggesting a direct relation between the physical properties
of these horizon shells and the recently proposed BMS supertranslation
hair of a black hole.

\newpage

\begin{small}
\tableofcontents
\end{small}
\vspace{.5cm}


\section{Introduction}

In this paper we revisit the theory of null shells in general relativity
\cite{BI}-\cite{poisson2}, with a particular emphasis on null shells
placed at horizons of black holes.

It seems to be a somewhat underappreciated fact that (in contrast
to what happens for spacelike or timelike shells) there can be
considerable freedom in ``soldering'' two geometries along a given null
hypersurface while maintaining the Israel junction condition
that the induced metric is continuous across the surface. 
Even though this was pointed out in \cite{BI}, and isolated
examples were already known in the literature before
(e.g.\ the Dray 't Hooft shell separating (or joining) two equal mass
Schwarzschild black holes along their horizon \cite{DT2}), there
seems to have been no systematic subsequent analysis of this phenomenon. 

We therefore begin with a systematic analysis of the conditions under
which non-trivial soldering transformations can exist, and we also
place these results into the general setting of Carrollian manifolds
and associated notions of symmetry groups, as recently formulated in
\cite{dg1, dg2}.

It follows from this analysis that the resulting soldering group is
infinite-dimensional when the induced metric on the null hypersurface
is invariant under translations along the null generators of the null
hypersurface. This condition is of course satisfied by Killing horizons of
stationary black holes, but also by Rindler horizons, and more generally
by other quasi-local notions of horizons such as non-expanding horizons
and isolated horizons (see e.g.\ \cite{ih1}, \cite{ih2} for reviews). 

In all these cases, we can generate an infinite number of physically distinct
shells on this hypersurface, and we will refer to these generically as
\textit{Horizon Shells}. These shells are parametrised by one (essentially)
arbitrary function on the horizon, which arises as an 
arbitrary coordinate transformation of the null coordinate $v$
on the hypersurface (with coordinates $(v,x^A)$),
\be
\label{igst}
v \ra F(v,x^A)
\ee
(extended to a suitable coordinate transformation off the shell
on one side of the shell).\footnote{Transformations of the form 
\eqref{igst} have arisen previously in the discussion of asymptotic
symmetries of Killing horizons in \cite{koga}. 
We thank Andy Strominger for bringing this reference to 
our attention.}
The resulting shells will in general carry a (null) matter
energy momentum tensor (composed of energy density, energy currents and
pressure), as well as impulsive gravitational waves travelling along the
shell, and we determine and analyse these in some detail. 

In particular, these horizon shells give rise to a rich \textit{classical}
structure at the horizon of a Schwarzschild black hole, and 
can be considered as significant generalisations of the
Dray 't Hooft null shell \cite{DT2}. As all of them are non-singular
on the shell (i.e.\ do not have any point particle singularities),
they also provide one with a wide array of smoothed out versions of
Dray 't Hooft impulsive gravitational waves \cite{DT1}. Given the recent
interest in these configurations in the context of
holography and scattering from black hole horizons (see e.g.\ \cite{StSh, Polchinski}), 
it is perhaps of some interest to have these new shell solutions at one's disposal.

A class of soldering transformations that may be of particular interest, 
especially in light of the observation in \cite{andy2} that black holes
must carry supertranslation hair, and the subsequent
Hawking - Perry - Strominger proposal
\cite{Hawking}, 
are the horizon analogues of BMS supertranslations at $\mathcal{I}^+$, of the form 
\be
\label{ibms}
v\ra v + T(x^A)\;\;.
\ee
In particular, we find that for shells generated by supertranslations of  
the Kruskal coordinate $V$, the conserved energy $E[T]$ of the resulting shell
is of the form \eqref{ET}
\be
E[T]= \frac{1}{8\pi} \int_{S^2}T
\ee
(there is also a simple, but slightly different, expression \eqref{efet} for the conserved
energy of shells generated
by supertranslations of the advanced coordinate $v$).
This expression bears a tantalising similarity to the standard expression for
BMS supertranslation charges at $\mathcal{I}^+$ (see e.g.\cite{wz,
bt1, andy1})  and to an analogous expression for a near-horizon BMS
supercharge proposed recently in \cite{dggp}, and therefore suggests
a direct relation between BMS supertranslation hair of a black hole
and properties of the horizon shell. 

Since (modulo isometries) there is a one-to-one correspondence between
soldering transformations and null shells, we can think of the shell as
a faithful bookkeeping device for BMS-like and more general soldering
transformations. We therefore propose to interpret the abstract BMS
charges in the context of horizon BMS transformations concretely as
conserved physical charges (the energy, say) of the horizon shell that is
(or could be) generated by the corresponding soldering transformation.

Since we were led to discover these structures from a systematic
analysis of soldering transformations, and not from attempting to transfer
structures from $\mathcal{I}^+$ to the horizon, there are some differences
in perspective. For instance, we extend the soldering transformation from
the horizon to a coordinate transformation inside the horizon rather
than to the exterior (as would have been more natural in terms of the
BMS perspective). The two procedures are of course completely equivalent.

Moreover, we require no notion of asymptotic symmetry but consider the
most general transformation maintaining only the continuity of the induced
metric. It is straightforward, however, to specialise our construction
to whatever notion of asymptotic symmetry at the horizon is the relevant one
in the case at hand.
In particular, it is possible that not all the soldering
transformations \eqref{igst} allowed by the general construction are
actually relevant in a specific context (like that of horizon hairs
created by physical perturbations of the black hole metric, say), but only
a specific subset of them, like the above supertranslations \eqref{ibms}.

Finally, although there are many discussions of the role of conformal
invariance in the physics of event horizons, we see no place for
conformal transformations that are non-trivial on the horizon from our
shell perspective.

\begin{center}
------------------------
\end{center}

In section 2 we analyse the soldering freedom in the Israel junction
condition, and in section 3 we phrase these results in the language of
Carrollian manifolds. Since we use the framework of continuous coordinates
(coordinates in which all of the components of the metric are continuous
across the shell, not only those of the induced metric) to derive the
physical properties of the shell, in section 4  we explain how to lift
soldering transformations off the shell in such a way that this continuity
condition is satisfied. In section 5, we then obtain the general expressions
for the energy-momentum tensor and impulsive gravitational wave components
on the shell and discuss the corresponding conservation laws. In section
6, we present in an elementary fashion our main example, namely the
general horizon shell on the horizon $U=0$ of the Kruskal geometry,
determine and analyse in detail its physical properties and
study various special cases. Section 7 concludes with a brief discussion
of the same example in Eddington-Finkelstein coordinates.

\section{Soldering Freedom and Horizon Shells}
\label{secsol1}

The general construction of null shells in general relativity
\cite{BI, BH, poisson1, poisson2} involves a
matching of two manifolds with boundary, ${\mathcal V^+}$ to the
future of a null hypersurface ${\mathcal N}^+$ and ${\mathcal V^-}$
to the past of ${\mathcal N}^-$ to each other across a common null
boundary ${\mathcal N}$. Each of the manifolds ${\mathcal V^\pm}$
and ${\mathcal N}$ respectively have independent coordinate charts
$x^\mu_\pm$ for $\mu=0,1,2,3$, and $y^a$ for $a=0,1,2$ and $\V^\pm$
carry metrics $g^\pm_{\mu\nu}$.

The basic requirement for the matching of the future and past geometries
across the null hypersurface $\N$ is the (Israel) \textit{junction condition}
that the induced metrics 
\be
g^\pm_{ab} = g^\pm_{\mu\nu}|_{\N^\pm} \frac{\del x_\pm^\mu}{\del y^a}\frac{\del x_\pm^\nu}{\del y^b}
\ee
on the future and past boundaries are isometric. It is common to write this condition as
\be
\label{junc1}
[g_{ab}]\equiv g_{ab}^+ - g_{ab}^- = 0\;\;,
\ee
expressing the statement that there is no jump in the induced metric 
across the shell.
This condition ensures on the one hand that the two boundaries can be
identified with the hypersurface ${\mathcal N}$ with a unique metric, and
on the other hand this construction leads to solutions of the Einstein
equations that can be interpreted as describing a shell of matter
and/or gravitational radiation separating (or joining) the two regions.  

The general formalism of \cite{BI}-\cite{poisson2} then shows how to derive the intrinsic
properties of the shell, independently of any coordinate choices. We will
return to this issue later on. First of all, however, we want to ask
and address the question whether there is any freedom in this procedure
to ``solder'' the two geometries together or if it is completely specified
by specifying the two geometries (satisfying the junction conditions)
on the two sides. 

For a generic hypersurface, the metric induced on the hypersurface will
be a function of the way the hypersurface is embedded in the surrounding
space-time. Thus, for generic hypersurfaces the interior and exterior
space-times are soldered together in an essentially unique way dictated
by the embedding (we will see this explicitly below, also for non-null
shells). However, it was already pointed out in \cite{BI} by way of
example that for certain null shells like the Killing horizons of static
black holes, there is considerable freedom in how the two geometries
are attached, allowing one to slide one of the manifolds independently
along the null (isometry) direction on $\N$ before soldering. We will
now investigate this question more sytematically.

In order to (significantly) simplify the analysis, as well as the
calculations for specific examples later on in this paper, we will now
introduce a preferred class
of coordinate systems $y^a$ on the shell $\N$, and also a corresponding class
of space-time coordinates $x^\alpha$ in a neighbourhood of $\N$ (details of this
rather standard construction can found e.g.\ in \cite{BI}-\cite{poisson2}). 
\begin{enumerate}
\item We introduce a coordinate system $y^a =(v,y^A)$ on $\N$ adapted to the 
fact that any null hypersurface is generated by null geodesics 
(the integral curves of any null normal
$n$ of $\N$), i.e.\ $v$ is a parameter along the null geodesics related to the choice of
null normal $n$ by $n=\del_v$, and the remaining
spatial coordinates $y^A$ are used to label the individual null geodesics. In these
coordinates the induced degenerate metric on $\N$ has the form
\be
g_{ab}n^b = g_{av}=0 \quad\Ra\quad g_{ab}(y) dy^a dy^b = g_{AB}(v,y^C) dy^A dy^B\;\;,
\ee
the conditions $g_{vv}=g_{vA}=0$ expressing the fact that $\del_v$ is null and normal to $\N$.
\item We introduce a coordinate system $x^\alpha$ in a 2-sided neighbourhood of the shell
$\N$ (e.g.\ via a null analogue of the construction of Gaussian normal coordinates)
such that in this coordinate system 
all the components $g^\pm_{\alpha\beta}$ of the metric, not just those
contributing to the induced metric, are continuous across $\N$,
\be
\label{junc2}
[g_{\alpha\beta}]=0\;\;.
\ee
\end{enumerate}
We should add here that typically such a coordinate system is only
used as an intermediate auxiliary device in the literature, 
and is generally considered to be somewhat (or grossly \cite{poisson1})
impractical for actual calculations. However, 
for the applications of the
formalism that we are interested in, the construction of such a coordinate
system in a small neighbourhood of the shell turns out to be completely straightforward,
since we start with a
space-time metric with no shell (or, equivalently, philosophical questions
aside, with an empty shell), which obviously already comes with its
continuous coordinate system.

Anyway, given such a coordinate system the null boundary hypersurface
$\N$ can then be described by an equation $\Phi(x)=0$ where $\Phi(x)$ is
a smooth function such that $\Phi>0$ to the future $\V^+$ of ${\mathcal
N}$, and $\Phi<0$ on $\V^-$. It is then natural and convenient to choose
one of the coordinates $x^\alpha$ to be proportional to $\Phi(x)$, so
we set $x^\alpha = (u,x^a)$, with 
\be
\Phi(x) = \lambda u 
\ee
(for some conveniently chosen constant $\lambda$), and thus $\N$ is simply
given by $u=0$. Finally, the two coordinate systems 
are linked by the choice that on $\N$ one has 
\be
x^a|_{\N}=y^a \;\;.
\ee
(and this is the only slightly non-standard but very natural and convenient choice 
that we make). Therefore our coordinates are 
\be
(x^\alpha) = (u,x^a) = (u,v,y^A)\;\;.
\ee
More on the choice of defining function $\Phi(x)$ 
and our conventions regarding its relation with the normal vector $n$ can be 
found in section \ref{secpps}.

Equipped with this, we can now return to the question raised above
regarding the uniqueness of the soldering procedure.
Since a geometry is determined by a metric up to coordinate
transformations, one way to approach this question is to enquire if there
is the freedom to perform coordinate transformations on one side, say on
$\V^+$, while maintaining the fundamental junction condition $[g_{ab}]=0$.\footnote{One could 
of course obtain the same results
by performing a coordinate transformation only on $\V^-$. These two different perspectives
can then clearly be related via a smooth global coordinate transformation without altering the
intrinsic properties of the shell.}
Since on $\N$ we
have identified the coordinates $x^a$
with the intrinsic coordinates $y^a = (v,y^A)$, this amounts to
asking under which coordinate transformations of the $y^a$ the induced
metric $g_{ab}$ remains invariant. Infinitesimally 
\be
L_Z g_{ab} = Z^c\del_c g_{ab} + (\del_a Z^{c})g_{cb} + (\del_b Z^c)g_{ac} = 0\;\;.
\ee
For a \textit{timelike} (or spacelike) shell, to which these considerations up to this
point would also apply, this is just the Killing equation, and therefore
this argument shows that (as is well known from other perspectives) in 
this case the soldering is unique up to isometries of the induced metric
(and these leave all physical quantities invariant). 

For a \textit{null} shell, with its degenerate metric, 
with $g_{av}=0$ and $g_{AB}$ non-degenerate,
the situation is potentially more interesting. 
In this case, the above equation becomes
\be
\label{lxg}
L_Z g_{ab} = Z^c\del_c g_{ab} + (\del_a Z^{C})g_{Cb} + (\del_b Z^C)g_{aC} = 0
\ee
for a vector field $Z= Z^c\del_c = Z^v \del_v + Z^C\del_C$ on $\N$. 
For $a=v$ or $b=v$ one finds the condition 
\be
\label{lzg1}
L_Z g_{av}=0 \quad\Ra\quad \del_v Z^A = 0 \;\;,
\ee
which rules out any $v$-dependent transformations of the spatial coordinates $y^A$.
From the spatial components of \eqref{lxg} one finds
\be
\label{lzg2}
L_Z g_{AB}=0 \quad\Ra\quad 
Z^v \del_v g_{AB} +  Z^C\del_C g_{AB} + (\del_A Z^{C})g_{CB} + (\del_B Z^C)g_{AC} = 0\;\;.
\ee
The corresponding group of allowed transformations does not depend on a particular choice of
normalisation of the null generators, and is thus an object intrinsically associated to the null
hypersurface and its metric. 
It is just the isometry group $\mathsf{Isom}(g_{ab})$ of the degenerate metric on $\N$, 
with Lie algebra 
\be
\mathfrak{isom}(\N,g_{ab}) = \{Z:\;L_Z g_{ab}=0\}\;\;.
\ee
We will now take a closer look at the solutions of \eqref{lzg2}. For a generic
$v$-dependence of $g_{AB}$, one  only has rigid (i.e.\ $v$-independent)
isometries of the metric $g_{AB}$, with $Z^v=0$, i.e.
\be
\text{generically:}\quad \mathsf{Isom}(\N,g_{ab}) = \mathsf{Isom}(g_{AB}) 
\ee
(and as in the case of timelike shells there is then an essentially unique soldering).

While there are some special cases in which a $v$-dependent metric can possess
non-trivial soldering transformations, as in the example of the Nutku-Penrose 
construction of impulsive gravitational waves on the light cone \cite{NP} (cf.\ 
also section 1.2 of \cite{BH} or the light cone example in \cite{dg1, dg2}), 
there is an especially interesting case that we will focus on 
here, in which the soldering group is not only non-trivial but actually
infinite dimensional. This happens 
when the metric is
independent of $v$, i.e.\ translation invariant along the null generators
of $\N$, 
\be
n^c\del_c g_{ab} = 0 \quad\LRa\quad \del_v g_{AB}=0\;\;.
\ee
In particular, this includes the \textit{Killing Horizons} of stationary black holes 
and \textit{Rindler Horizons} (with their boost Killing vector).
More generally, since our considerations involve only $\N$
itself, this condition, which implies that the null congruence on $\N$ has
zero expansion and shear, is satisfied by arbitrary \textit{Non-expanding Horizons}, a precursor
to \textit{(Weakly) Isolated Horizons} (see e.g.\ \cite{ih1}, \cite{ih2} for reviews).
In the current context of null shells, we will therefore refer to null 
hypersurfaces with a metric satisfying $\del_v g_{ab}=0$ or
$\del_v g_{AB}=0$ as \textit{Horizon Shells}.

In this case the component $Z^v$ is completely unconstrained, and corresponds
to the freedom to perform arbitrary coordinate transformations of $v$
on $\N$,
\be
\label{nu1}
\del_v g_{AB} = 0 \quad\Ra\quad v\ra  F(v,y^A)\quad\text{allowed}
\ee
(here and in the following it is understood that the function $F$ is such that 
$(v,y^A)\ra (F, y^A)$ is a legitimate orientation preserving coordinate transformation, 
i.e.\ that $F$ satisfies $\del_v F > 0$). 
Because of the presence of an arbitrary function $F$ in the 
coordinate transformation, the isometry group 
$\mathsf{Isom}(\N,g_{ab})$ is infinite dimensional, and factorises as the
semi-direct product
\be
\text{Horizon Shells}:\quad \mathsf{Isom}(\N,g_{ab}) =  
\mathsf{Isom}(g_{AB}) \ltimes \mathsf{Sol}(\N)\;\;,
\ee
where the infinite dimensional soldering group $\mathsf{Sol}(\N)$
(of non-trivial soldering transformations) is 
the group of coordinate transformations \eqref{nu1}, 
\be
\label{nu2}
\mathsf{Sol}(\N,g) = \{v \ra F(v,y^A)\}\;\;.
\ee
As a consequence there are e.g.\ an infinite number of ways to glue two 
black holes geometries together (the inside horizon region of one to the 
exterior region of the other) provided only that the Israel junction 
condition is satisfied. In sections \ref{seckruskal} and \ref{secef} 
we will analyse in detail the simplest example exhibiting this 
phenomenon, namely the soldering of two 
equal mass Schwarzschild metrics along their 
horizon.

Let us also analyse the effect of the transformations generated by $Z$ 
on the normal null generator $n^a$ of $\N$ for a general null shell. Since the metric is invariant
by the condition \eqref{lxg}, 
and $n^a$ spans the kernel of the metric, it is clear a priori that $L_Z$ preserves the 
direction of $n^a$, $L_Zn^a \sim n^a$, 
\be
\begin{aligned}
g_{ab}n^b = 0 &\quad\Ra\quad 0 = 
L_Z(g_{ab}n^b) = 
(L_Zg_{ab})n^b + g_{ab}L_Z n^b = 
g_{ab}L_Z n^b \\
&\quad\Ra\quad L_Zn^a \sim n^a \;\;.
\end{aligned}
\ee
This can also be seen explicitly from $n=\del_v$, i.e.\ $n^a = \d^a_v$, and \eqref{lzg1}, 
\be
L_Zn^a = Z^b \del_b n^a - n^b \del_b Z^a = -\del_v Z^a = - (\del_v Z^v) \d^a_v = -(\del_v Z^v) n^a\;\;.
\ee
If one further restricts the transformations to those that strictly preserve
the normal $n^a$, then the allowed coordinate transformations on the shell
are restricted to $\del_v Z^v=0$. In the case of horizon shells, this 
restricts the soldering transformations \eqref{nu1} to
\be
\label{bms1}
L_Zn^a = 0 \quad\Ra\quad \del_v Z^v = 0 \quad\Ra\quad 
v\ra  v + F(y^A)\quad\text{allowed}\;\;.
\ee
Thus even when there is a preferred (Killing, say) null generator of the horizon 
shell, the soldering group preserving this structure is
still infinite dimensional.

\section{Soldering Group, Carrollian Manifolds and BMS Transformations}
\label{secbms}

Interestingly the above considerations, motivated by the question of the freedom in 
soldering null shells, are closely related to recent investigations of the 
(ultra-relativistic) Carroll group and various other symmetry groups of Carrollian manifolds 
and Carrollian structures and their relation with BMS supertranslation 
symmetries (cf.\ in particular \cite{dg1, dg2}).

A \textit{Carrollian manifold} is by definition (we adopt the ``weak''
definition of \cite{dg1, dg2}) a manifold equipped with a degenerate
non-negative metric whose kernel is everywhere 1-dimensional (and is thus
spanned by a nowhere vanishing vector field). Clearly, according to this
definition any null hypersurface of a Lorentzian (pseudo-Riemannian) 
space-time defines a Carrollian manifold, the Carroll structure being
encoded in the triplet $(\N, g_{ab}, n^a) \equiv (\N,g,n)$.\footnote{For the converse question,
how to embed a given Carrollian manifold as a null hypersurface into an
ambient space-time, see \cite{xbkm, jelle}.} 

Given a Carrollian structure, one can then analyse various notions of 
symmetry groups preserving (in a suitable sense) such a structure. In 
particular, the group of transformations preserving both $g_{ab}$ and
$n^a$ could be called the isometry group $\mathsf{Isom}(\N,g,n)$. Its
Lie algebra is generated by the vector fields $Z$ on $\N$ satisfying
\be
\mathfrak{isom}(\N,g,n) = \{Z:\;L_Zg = L_Zn = 0\}\;\;.
\ee
As we have seen, in the case of horizon shells (as defined above)
this isometry
group is infinite dimensional, due to the presence of the transformations
\eqref{bms1}. By extrapolation from the corresponding terminology for 
future null infinity $\mathcal{I}^+$, which is a natural example of a Carrollian
manifold, these transformations are called 
\be
\text{BMS supertranslations}:\quad 
v\ra  v + F(y^A)\;\;,
\ee
and form an infinite dimensional Abelian subgroup of the isometry group of the Carrollian
structure on such a null surface. 

Note that, via the identification $n=\del_v$, this definition of horizon shell 
BMS transformations depends on the choice of null normal. For example, unlike  
at $\mathcal{I}^+$, at the horizon of a static black hole supertranslations
of the Killing parameter (Eddington-Finkelstein advanced $v$) are not the 
same as supertranslations of the affine parameter (Kruskal $V$).

In passing we also note that for horizon shells $(\N,g,n)$ also defines a Carrollian structure
in the strong sense of \cite{dg1, dg2}, which requires the existence of a
symmetric affine connection $\nabla$ compatible with $(\N,g,n)$, i.e. 
\be
\nabla_c g_{ab} = 0\quad,\quad \nabla_c n^a = 0\;\;.
\ee
Indeed, it is easy to see that any symmetric connection $\Gamma^a_{\;bc}$ with 
\be
\Gamma^A_{\;BC}=\Gamma^A_{\;BC}(g)\quad,\quad \Gamma^a_{\;bv} = 0 \quad,\quad \Gamma^v_{\;AB}
\quad\text{arbitrary}
\ee
($\Gamma^A_{\;BC}(g)$ are the components of the Levi-Civita connection of the spatial metric
$g_{AB}$) satisfies this condition. As the existence of such a connection does not 
appear to play a significant role in the context we are interested in, there is also no
reason to restrict the soldering transformations to the group of transformations 
that preserve the connection (which would be finite-dimensional). 

In the Carrollian setting, the more general allowed soldering transformations \eqref{nu1} 
appear as a subgroup of what (again by extrapolation from $\mathcal{I}^+$) are called
\textit{Newman-Unti transformations}. Due to its $\mathcal{I}^+$ pedigree, the Newman-Unti
group also contains conformal isometries. In the present context 
of the soldering of horizon shells, conformal isometries are not
allowed (as they fail to satisfy \eqref{lzg2}), and the relevant 
group is the group $\mathsf{Sol}(N,g)$ \eqref{nu2} of soldering 
transformations.

\section{Off-Horizon Shell Extension of the Soldering Transformations}
\label{secsol2}

In keeping with the framework we have adopted for discussing shells, 
we now need to understand how to
extend the soldering transformations off the shell $\N$ in such a way that 
the entire metric remains continuous across the shell, 
\be
\label{gcont}
[g_{\alpha\beta}]=0\;\;.
\ee
We thus asume that we have a ``seed'' metric which satisfies \eqref{gcont} (in the example
of section \ref{seckruskal} this is simply the Schwarzschild metric in Kruskal
coordinates, with an empty shell on the horizon $U=0$).

We now want to lift the soldering transformation generated by $Z=Z^v\del_v$ on $\N$
to a coordinate transformation in an infinitesimal neighbourhood of one side of the shell, 
say $\V^+$, in such a way that the continuity of the metric \eqref{gcont} is 
maintained. I.e.\ we extend the vector field off $\N$ as
\be
\label{soldtrans}
Z_+= Z^v\del_v + u z^\alpha\del_\alpha 
\ee
and impose the requirement that 
\be
L_{Z_+}g_{\alpha\beta}|_{\N}=0 \;\;.
\ee
Because on $\N$ the vector field generates a soldering transformation, the conditions
\be
L_{Z_+}g_{ab}|_{\N}=0 
\ee
are identically satisfied. Indeed, since 
\be
Z_+|_{\N} = Z^v \del_v \quad, \quad \del_aZ_+|_{\N} = (\del_a Z^v)\del_v\;\;,
\ee
one has 
\be
\begin{aligned}
L_{Z_+}g_{ab}|_{\N} &= 
\left(Z^v \del_v g_{ab} + \del_a Z^\alpha g_{\alpha b} + 
\del_b Z^\alpha g_{a\alpha}\right)|_{\N}\\
&=
\left(Z^v \del_v g_{ab} + \del_a Z^v g_{v b} + \del_b Z^v g_{av}\right)|_{\N} = 0
\end{aligned}
\ee
because on the null horizon shell $\del_v g_{ab}= g_{av}=0$.
It thus remains to impose the conditions
\be
L_{Z_+} g_{u\beta}|_{\N}=0\;\;.
\ee
These are linear equations for the coefficients $z^\alpha$,
\be
\begin{aligned}
\beta = u:\quad& Z^v \del_v g_{uu} + 2 z^\alpha g_{\alpha u} = 0\\
\beta = v:\quad& Z^v \del_v g_{uv} + (z^u + (\del_v Z^v)) g_{uv} = 0\\
\beta = B:\quad& Z^v \del_v g_{uB} + z^\alpha g_{\alpha B} + (\del_B Z^v) g_{uv} = 0
\end{aligned}
\ee
(here the restriction to $\N$ is implied).

In particular, if we now restrict to metrics with 
\be
g_{uu}= g_{uA}=0\quad,\quad (\del_v g_{uv})|_{\N} = \del_v (g_{uv}|_{\N}) = 0\;\;,
\ee
(typical examples would be spherically symmetric metrics in null or
double-null coordinates like the Schwarzschild metric in
Eddington-Finkelstein or Kruskal coordinates),  
the solutions to the above equations are
\be
\label{infZ}
z^v = 0\quad,\quad z^u = -\del_v Z^v \quad,\quad z^A = - g_{uv} g^{AB}\del_B Z^v \;\;.
\ee
E.g.\ for the Schwarzschild metric in Kruskal coordinates (with $u=U,v=V$) one has 
\be
\label{infK}
z^V=0\quad,\quad z^U = -\del_VZ^V\quad,\quad z^A = (2/e) \sigma^{AB}\del_B Z^V\;\;,
\ee
with $\sigma_{AB}$ the standard metric on the unit 2-sphere. 
Defining the function $\omega(v,\theta,\phi)$ by 
\be
Z^V = V \omega(V,\theta,\phi)\;\;,
\ee
the generator \eqref{soldtrans} of the off-shell extension of the soldering 
transformation can be rewritten in the suggestive form
\be
Z_+ = \omega\left(V\partial_V - U\partial_U\right) 
+ U (z^A\partial_A -V(\del_V\omega))\del_U
\ee
of a ``local Killing transformation'' (with respect to the Kruskal Killing vector
$\sim V\del_V - U\del_U$ and with local coefficient $\omega(V,\theta,\phi)$),
as was first found in the discussion of the asymptotic symmetries of Killing horizons
in \cite{koga}. That the soldering transformation can be written as a local 
Killing transformation
is even more transparent in 
Eddington-Finkelstein coordinates, with Killing vector $\del_v$, where
\eqref{soldtrans} has this form on the nose.

In principle this can now be exponentiated to find the exact coordinate
transformation to linear order in $u$. In practice, however, this is a
bit tedious, and in the following we will obtain this transformation
directly, starting from the ansatz
\be
\label{ansatz}
v_+ = F(x^a) + u A(x^a)\quad,\quad x^A_+= x^A + uB^A(x^a) \quad,\quad u_+ = u C(x^a)
\ee
(the linear term $uA$ in $v_+$ will be generated by exponentiation of
\eqref{infZ}),
and demanding continuity of the metric, $[g_{\alpha\beta}]=0$. 
Continuity of $g_{u\alpha}$ then determines the functions 
$B^\alpha = (A,C,B^A)$.

\section{From Soldering to the Physical Properties of the Shell}
\label{secpps}

Assuming that we have successfully implemented the steps of finding 
a non-trivial soldering transformation and its off-shell lift, we
can now determine (pretty much read off) the physical properties of
the shell. 

We just need to be slightly more specific about our 
conventions regarding the choice of normal vector. Recall 
from section \ref{secsol1} that the shell is given
by the equation $\Phi(x)=\lambda u = 0$.
A null normal (and tangent) to ${\mathcal N}$ is then given by
\be
\label{nalpha}
n_\alpha = -\partial_\alpha\Phi|_{\N} = -\lambda \del_\alpha u|_{\N}\;\;.
\ee
Here the sign is chosen such that $n^\alpha$ is future pointing when $\Phi$ increases towards
the future, and 
we absorb the freedom to multiply $n^\alpha$ by an arbitrary non-vanishing function on 
$\mathcal{N}$ into the freedom in the choice of $\Phi$. Since on $\N$ we had already chosen
$n=\del_v$, this correlates the choice of $\Phi$ with 
the choice of coordinate $v$, 
\be 
n^\alpha\del_\alpha = n^a \del_a = \del_v\;\;.
\ee
As the metric is continuous across the shell in the chosen coordinates,
so are all its tangential derivatives. 
In order to be able to take derivatives in the direction transverse
to the shell, we need to
introduce an auxiliary vector field $N$ that is transverse to $\N$,
i.e.\ $n.N\neq 0$, and continuous across $\N$, $[N]=0$. 
A convenient choice is
\be
\label{Nchoice}
N = \lambda^{-1}\del_u \quad\Ra\quad N^\alpha n_\alpha =-1\;\;.
\ee
It then turns out that the complete information
about the intrinsic physical properties of the shell is encoded in the
first transvese derivative of the tangential components of the metric
(the choice to make the non-tangential components $g_{\alpha u}$ continuous
was just a gauge choice and has no influence on the physics).
Therefore
the basic shell-intrinsic tensor that contains all information about the 
properties of the shell is 
\be
\label{gamma2}
\gamma_{ab} = N^\alpha[\partial_\alpha g_{ab}] = 
\lambda^{-1}[\partial_u g_{ab}] = \trac{1}{2}[L_N g_{ab}]\;\;.
\ee
A simple way to determine the $\gamma_{ab}$ is to 
expand the tangential components of the 
interior and exterior metrics to linear order in $u$, 
\be
\dd s^2 = (g^{(0)}_{ab} + u g^{(1)\pm}_{ab} + \ldots)\dd x^a\dd x^b\;\;,
\ee
and to then read off the $\gamma_{ab}$ from 
\be
\label{gamma3}
\gamma_{ab} = \lambda^{-1}[g_{ab}^{(1)}] \;\;.
\ee
Starting from the soldering transformation $v \ra F(v,y^A)$, the $\gamma_{ab}$ are then
given explicitly in terms of $F$ and its 1st and 2nd derivatives, 
\be
\gamma_{ab} = \gamma_{ab}(F, \del_a F, \del_a\del_b F)\;\;.
\ee
In terms of the $\gamma_{ab}$, the intrinsic energy-momentum tensor of the shell is in turn 
given by
\be
16\pi S^{ab} = -\gamma^* n^an^b - \gamma^\dagger g^{ab}_* + 2\gamma^{(a}n^{b)}
\ee
where $g^{ab}_*=\delta^a_{\;A}\delta^b_{\;B} g^{AB}$ is the inverse of the spatial (non-degenerate) part of 
the hypersurface metric and
\be
\label{gammac}
\gamma^* = g_*^{ab}\gamma_{ab} = g^{AB}\gamma_{AB}\quad,\quad \gamma_a = \gamma_{ab}n^b
\quad,\quad
\gamma^a = g_*^{ab}\gamma_b
\quad,\quad \gamma^\dagger = \gamma_{ab} n^a n^b = \gamma_a n^a\;\;.
\ee
The components of $S^{ab}$ can be interpreted as surface energy density $\mu$,
pressure $p$ and energy currents $j^A$, where 
\be
\mu = -\frac{1}{16\pi}\gamma^*\quad,\quad j^A = +\frac1{16\pi}\gamma^A\quad,\quad
p = -\frac1{16\pi}\gamma^\dagger\;\;.
\ee
In particular, the pressure is related to the jump in the surface gravity or 
inaffinity $\kappa$, defined by 
\be
n^\beta \nabla_\beta n^\alpha = \kappa n^\alpha \;\;.
\ee
Indeed, taking the scalar product with the transverse vector $N$ and using $N.n=-1$ and 
the definition of 
$\gamma_{ab}$ \eqref{gamma2}, one finds
\be
\kappa = - 
n^\beta N_\alpha \nabla_\beta n^\alpha = 
(\nabla_\beta N_\alpha) n^\beta n^\alpha = \trac{1}{2}(L_Ng_{\alpha\beta})n^\alpha n^\beta
= \trac{1}{2}(L_Ng_{ab})n^a n^b\;\;,
\ee
and therefore
\be
[\kappa] = \trac{1}{2}\gamma_{ab} n^a n^b = \trac{1}{2}\gamma^\dagger \;\;.
\ee
Note that there is an ambiguity in the operational interpretation of the components of 
\be
S^{ab} = \mu n^a n^b + j^a n^b + j^b n^a + p g_*^{ab}
\ee
due to absence of a rest-frame on the null shell
${\mathcal N}$. As a consequence, different observers upon crossing the shell will 
measure rescalings of $\mu$, $p$ and $j^A$ as described in detail in section 3.11 of
\cite{poisson2}.

For a general null shell the surface energy-momentum tensor has only 
these 4 independent components (in contrast to a timelike shell, which has
6), while $\gamma_{ab}$ has 6 components. Indeed it is evident from 
the expressions \eqref{gammac} that the transverse traceless components
$\hat{\gamma}_{ab}$ of $\gamma_{ab}$, characterised by
\be
\hat{\gamma}_{ab}n^b = 0\quad,\quad g_*^{ab}\hat{\gamma}_{ab}=0\;\;,
\ee
do not contribute to the matter content on the shell encoded in 
$S^{ab}$. Instead, these components, which can be extracted
from $\gamma_{ab}$ according to
\be
\label{gammahat}
\hat{\gamma}_{ab} = \gamma_{ab} -\frac12 \gamma^* g_{ab} + 2
\gamma_{(a}N_{b)} + (N_aN_b-\frac{1}{2}N.N g_{ab})\gamma^\dagger \;\;,
\ee
contribute to the Weyl tensor on the shell and 
describe the 2 polarisation states of an impulsive gravitational wave
travelling along the shell.
For more details on this gravitational wave component 
in general see e.g.\ section 2.3 of \cite{BH}.

Finally, we note that the shell
energy-momentum tensor $S^{ab}$ satisfies certain conservation laws derived
in \cite{BI} (their derivation requires some care, due to the degeneracy
of the metric, and because $S^{ab}$ is only defined on the shell).
In the absence of bulk matter these are 
\be
\label{conslaw}
N_a\left(\partial_b + \tilde{\Gamma}_b\right)S^{ab} - S^{ab}\tilde{\mathcal K}_{ab} = 0
\ee
and
\be
\label{3dcons}
S_{a;b}^b = \left(\partial_b + \tilde{\Gamma}_b\right)S_a^b - 
\frac12 S^{bc}\partial_ag_{bc}=0\;\;.
\ee
Here
\be
S^b_a \equiv g_{ac}S^{cb} = \frac{1}{16\pi}\left(n^b\gamma_a - \d^b_a \gamma^\dagger\right)\;\;,
\ee
a tilde over a quantity denotes an average value of the quantities from the two sides of the shell, 
$\tilde{\Gamma}_b$ is the average of the null surface counterpart of the contracted
Christoffel symbol, defined (in the coordinates that we have used here) as 
\be
\Gamma^\pm_b = \nabla^\pm_\mu \d^\mu_b = \Gamma^{\pm\mu}_{\;\mu b}\;\;,
\ee
and finally $\mathcal{K}_{ab}$ is the ``transvese extrinsic curvature'', 
satisfying $\gamma_{ab}=2[\mathcal{K}_{ab}]$ and given in our coordinates simply by
\be
\mathcal{K}^\pm_{ab} = (2\lambda)^{-1} g^{\pm (1)}_{ab}\;\;.
\ee
It may be a bit puzzling that one obtains 4 equations for $S^{ab}$. 
We will see that \eqref{conslaw} and the spatial components of \eqref{3dcons} give rise 
to the 3 expected conservation laws for $S^{ab}$. The $v$-component of \eqref{3dcons} (equivalently, 
the contraction of \eqref{3dcons} with $n^a$), however, 
is simply a geometric identity which (as such) is identically satisfied.
Temporarily including also bulk matter, this contracted equation reads
\be
\label{rayn1}
n^a S_{a;b}^b = [T_{\alpha\beta}n^\alpha n^\beta] = [T_{ab}n^a n^b]\;\;.
\ee
Noting that $n^a S_a^b = 0$, we find that the left-hand side is simply
\be
n^a S_{a;b}^b = 
- \frac12 n^a S^{bc}\partial_ag_{bc}=
- \frac12 S^{BC}\partial_vg_{BC}= -\frac12 p g^{BC}\del_v g_{BC} = \frac{1}{8\pi} 
[\kappa]\theta
\ee
where $\theta$ is the expansion of the null congruence on $\N$. On the other hand, the 
Raychaudhuri equation for this congruence reads (upon using the Einstein equations)
\be
\del_v\theta + \trac{1}{2}\theta^2 + \sigma^{ab}\sigma_{ab} = \kappa\theta - 
8\pi T_{ab}n^a n^b
\ee
(with $\sigma_{ab}$ the spatial shear tensor).
Since the left-hand side of this equation depends only on the intrinsic geometry of $\N$, 
its jump is zero, and therefore one finds \cite{poisson2}
\be
[\kappa]\theta = 8\pi [T_{ab}n^a n^b]
\ee
as a geometric identity, and comparison with the above shows that this is precisely 
the content
of \eqref{rayn1}. In the context of horizon shells, with no bulk matter, this equation is 
identically satisfied (without any constraint on the pressure) because the expansion 
$\theta = 0$.
The remaining 3 equations can be seen to be satisfied for the $S^{ab}$ that
we derive in the examples.\footnote{We should add that the fact that we obtain 4 
valid equations 
without any restrictions on the ``type'' of the null hypersurface appears to contradict
remarks in section 3.5 of \cite{BH} that this should not be possible.}

\section{Schwarzschild Horizon Shell in Kruskal coordinates}
\label{seckruskal}

To illustrate the above construction, we will now look in detail
at the Schwarzschild (Kruskal) horizon shell. Thus we start with the
Kruskal metric without a shell as our seed metric, and then we generate
non-trivial shells on the horizon between two equal mass black hole
metrics via soldering transformations and determine the physical
properties of the shell.

\subsection{General Construction}

In Kruskal coordinates the Schwarzschild metric is
\be
\dd s^2 = - 2 G(r)\dd U\dd V + r^2\dd \Omega^2\;\;,
\ee
with
\be
G(r) = \frac{16m^3}{r}\ex{-r/2m}\quad,\quad
UV = - (\frac{r}{2m}-1)\ex{r/2m}\;\;.
\ee
We choose the horizon shell $\N$ to be the Kruskal horizon $U=0$.
As discussed in the previous sections, we then have a large soldering freedom
\be
\label{Ksol}
V\ra F(V,\theta,\phi) 
\ee
as a consequence of the fact that the null generators are orbits of an
intrinsic isometry of the induced hypersurface metric, but we can 
also quickly rederive this from scratch here. To that end, we introduce
coordinates $(U_\pm,V_\pm,\theta_\pm,\phi_\pm)$ on $\V^+$ ($U\geq 0)$ 
and $\V^-$ ($U\leq 0)$ respectively, so that we have
\be
ds_{\pm}^2 = 
- 2 G(r_\pm)\dd U_\pm\dd V_\pm + r_\pm^2 (d\theta_{\pm}^2 + \sin^2\theta_{\pm}d\phi_{\pm}^2)
\quad\text{on}\;\;\V^\pm\;\;.
\ee
Therefore, the metric induced on the horizon $\N$ at $U=0 \ra r=2m$ is
\be
ds_{\pm}^2|_{\N} = 
4m^2 (d\theta_{\pm}^2 + \sin^2\theta_{\pm}d\phi_{\pm}^2)\;\;. 
\ee
This induced metric is continuous across the shell (i.e.\ satisfies the
Israel junction condition \eqref{junc1}) provided that we choose 
\be
\theta_{+}|_{\N} = \theta_{-}|_{\N}\quad,\quad \phi_{+}|_{\N} = \phi_{-}|_{\N}
\ee
(up to isometry rotations).
However, as this metric does not depend on the coordinates $V_\pm$ (Killing horizon), 
we are free to choose any relation 
\be
V_+|_{\N}=F(V_-,\theta_-,\phi_-)
\ee
between the coordinates $V_\pm$ while maintaining the junction 
condition.\footnote{As already mentioned in section \ref{secsol1}, 
we could of course equivalently perform this transformation on
the exterior coordinate $V_-$ instead of ``hiding'' it behind the horizon.}
This is precisely the soldering freedom \eqref{Ksol} mentioned above. 

As continuous coordinates we now choose the Kruskal coordinates
\be
x^\alpha = (U,x^a) = (U=U_-,V=V_-,\theta=\theta_-,\phi=\phi_-)\;\;,
\ee
and the coordinates on the shell are then (naturally)
taken to be $y^a = x^a = (V,\theta,\phi)$.
An obvious
candidate for the defining function $\Phi(x)$ with $\N=\{x:\;\Phi(x)=0\}$ is
\be
\Phi(x) = \Phi(U) = \lambda U
\ee
with a constant $\lambda$ to be suitably chosen. As explained in section 
\ref{secpps}, this gives rise to a corresponding choice of null normal
\eqref{nalpha}
\be
n_\alpha = -\del_\alpha\Phi|_{\N} = -\lambda \partial_\alpha U|_{\N} 
\quad\Ra\quad n = n^a \del_a  = (\lambda/G(2m))\del_V\;\;.
\ee
In order to match with our choice of coordinates on $\N$, we choose 
$\lambda = G(2m)$, and thus $n$ and the transverse null vector field 
$N$ with $n.N=-1$ \eqref{Nchoice} are
\be
n=\partial_V \quad,\quad N = \frac{1}{G(2m)}\partial_U\;\;.
\ee
In order to determine the physical properties of the shell generated
by the above soldering transformation, 
we now extend the soldering transformation off the shell to
a small neighbourhood of $\N$ in $\V^+$ such that all the components
$g_{\alpha\beta}$ of the metric are continuous across the shell, not
just those of the induced metric. 

To the order in $U$ necessary for the calculation 
of the energy-momentum tensor of the shell, 
the off-shell extension of the soldering transformation 
has the general form \eqref{ansatz}
\be
V_+ = F(V,x^A) + UA(V,x^A)\quad,\quad x_+^A = x^A + U B^A(V,x^A)\quad,\quad U_+ = UC(V,x^A)
\ee
(with $x^A = (\theta,\phi)$).
Requiring continuity of the metric in Kruskal coordinates, specifically of the 
components $g_{U\alpha}$, determines 
\be
\label{ABC}
C = \frac{1}{F_V}\quad,\quad B^A = \frac{2}{e}\frac{\sigma^{AB}F_B}{F_V}\quad,\quad A = 
\frac{e}{4} F_V \sigma_{AB}B^A B^B\;\;,
\ee
with $\sigma_{AB}$ the components of the metric on the unit 2-sphere, and 
$F_V=\del_VF, F_B=\del_BF$.

Infinitesimally, this coordinate transformation reduces precisely to
the transformation given in \eqref{infK}, with  $A$ being generated as
a higher order correction by exponentiation of \eqref{infK}.
One can easily show that $A$ can be
set to zero by adding higher order $U^2$-terms to the transformations
of the remaining coordinates (which has no effect on the properties
of the shell itself).\footnote{This is an alternative way to understand the
``gauge invariance'' of the space-time components $S^{\mu\nu}$ of the shell 
energy momentum tensor under
certain transformations of the space-time $\gamma_{\mu\nu}$ discussed in
\cite{BI}.} We can and will therefore set $A$ to zero for the calculation
below.

To calculate the energy-momentum tensor, we 
follow the procedure described in section \ref{secpps} and 
determine the $\gamma_{ab}$ from the first-order expansion of the metric 
\eqref{gamma3}
\be
\gamma_{ab} = \frac{1}{G(2m)} [g_{ab}^{(1)}] \;\;.
\ee
While we only need the leading order term in $G(r)$, 
$G(2m) = 8m^2/e$, 
we need to expand $r^2$ and $\sin^2 \theta_+$ to linear order in $U$.
On $\V_-$ we just need
\be
r(UV)^2 = 4m^2 - \frac{8m^2}{e}UV + \cdots \;\;, 
\ee
and on $\V_+$ we have
\be
\begin{aligned}
r(U_+V_+)^2 &= 4m^2 - \frac{8m^2}{e}U_+V_+  + \cdots = 4m^2 - \frac{8m^2}{e}\frac{UF}{F_V} + \cdots \\
\sin^2\theta_+ &= \sin^2\theta + 2UB^\theta\sin\theta\cos\theta +\cdots
\end{aligned}
\ee
Putting everything together we obtain
\be
\begin{aligned}
g^{(1)-}_{ab}\dd x^a\dd x^b &= -(8m^2/e)V\sigma_{AB}dx^A dx^B\\
g^{(1)+}_{ab}\dd x^a\dd x^b &= 8m^2 \left(-(2/e)dC\;dF + 
\sigma_{AB}dx^A (dB^B - (F/eF_V) dx^B) + \sin\theta\cos\theta
 B^\theta d\phi^2 \right)
\end{aligned}
\ee
Using the explicit expressions \eqref{ABC}, one then finds
\be
\label{anglegammas}
\begin{aligned}
\gamma_{Va} &= 2\frac{\partial_V\partial_aF}{F_V} = 2 \del_a \log F_V\\
\gamma_{AB} &= 2\left(\frac{\nabla^{(2)}_A\partial_B F}{F_V} 
- \frac12 \sigma_{AB}(\frac{F}{F_V} - V)\right)\;\;,
\end{aligned}
\ee
where $\nabla^{(2)}$ is the 2-dimensional covariant derivative associated with $\sigma_{AB}$. 
The explicit expressions for the energy density, surface currents and 
pressure are then
\be
\begin{aligned}
\label{Kmujp}
\mu &= -\frac1{32m^2\pi F_V}\left(\Delta^{(2)}F - F + VF_V\right)\\
j^A &= \frac1{32m^2\pi}\sigma^{AB}\frac{F_{BV}}{F_V}\\
p &= -\frac1{8\pi}\frac{F_{VV}}{F_V}\;\;.
\end{aligned}
\ee
These quantities satisfy the conservation laws \eqref{conslaw} and \eqref{3dcons}.
We had already observed that the null component of \eqref{3dcons} is identically 
satisifed. The conservation law \eqref{conslaw} can be shown to be
\be
\label{Kconslaw}
(\partial_V + \tilde{\kappa})\mu + (\nabla^{(2)}_A +\frac12\gamma_{AV})j^A 
+\frac14 (g^{AB}\gamma_{AB} -4V)p =0\;\;.
\ee
Here
$\tilde{\kappa}$ is the average value of the surface gravity (inaffinity) on 
the two sides. Since $V$ is affine on $\V^-$, one has 
\be
\tilde{\kappa} = \trac{1}{2}(\kappa^+ + \kappa^-) = \trac{1}{2}\kappa^+\;\;.
\ee
Then it is straightforward to check that \eqref{Kconslaw} is satisfied by the quantities in 
\eqref{Kmujp}. 

The spatial (angular) components of \eqref{3dcons}
become
\be
\label{Kangle}
\del_V\gamma_{AV} = \del_A \gamma_{VV} \quad\LRa\quad \del_V j_A + \del_A p = 0 \;\;,
\ee
which are obviously also identically satisfied by virtue of \eqref{anglegammas}. Note
also that the (gradient) currents have the additional property 
\be
\del_B j_A - \del_A j_B = 0\;\;.
\ee

Finally, the two remaining components $\hat{\gamma}_{ab}$ of 
$\gamma_{ab}$ \eqref{gammahat} give the 2 polarisations of an
impulsive gravitational wave travelling along the shell, and we find in the above
example that they are
\be
\label{gammahatK}
\begin{aligned}
\hat{\gamma}_{\theta\phi} &= \gamma_{\theta\phi} = 2\frac{\nabla^{(2)}_\theta\partial_\phi F}{F_V}\\
\hat{\gamma}_{\theta\theta} &=  -\frac1{\sin^2\theta}
\hat{\gamma}_{\phi\phi} = \frac12\left(\gamma_{\theta\theta} 
- \frac1{\sin^2\theta}\gamma_{\phi\phi}\right) = \frac{2}{F_V}
\left(\nabla^{(2)}_\theta\partial_\theta F -
\frac1{\sin^2\theta}\nabla^{(2)}_\phi\partial_\phi F\right)
\end{aligned}
\ee

\subsection{Special Cases}
\label{subk}

In order to get acquainted with these shells and their properties, we now specialise
the above general construction in various ways.

\begin{itemize}
\item Dray 't Hooft Shell

The simplest example is the Dray 't Hooft shell \cite{DT2} which
one obtains from a constant shift $V_+ = V + b$, with
\be
\mu = \frac{b}{8\pi(2m)^2}\quad,\quad j_A = p = 0\;\;.
\ee
As $\gamma_{AB}$ is pure trace for this constant shift, there is also
no accompanying gravitational wave contribution in this case. Thus all
the horizon shells we are considering that are generated by more general
soldering transformations can be regarded as a broad class of
generalisations of the Dray 't Hooft shell.

\item Zero Pressure: $p=0$

From $p\sim \gamma_{VV}\sim F_{VV}$ it follows that
\be
p = 0 \quad\Ra\quad F(V,\theta,\phi) = A(\theta,\phi)V + B(\theta,\phi)\;\;.
\ee
This result, an angle-dependent affine transformation of $V$, 
can also be understood from the fact that these are precisely
the transformations that leave the inaffinity (surface gravity) $\kappa$ 
invariant, so that $p \sim [\kappa] = 0$.

\item Zero Pressure and Vanishing Currents: $p=j_A = 0$

Since $j_A \sim \gamma_{VA}\sim F_{VA}$, we next deduce
\be
j_A = 0 \quad\Ra\quad F(v,\theta,\phi) = aV + B(\theta,\phi)\;\;.
\ee

\item BMS Supertranslations of $V$

From this we learn that, up to the irrelevant rescaling $V\ra aV$ (which is part
of the isometry $(V,U)\ra (aV,a^{-1}U)$ of the Kruskal metric), BMS supertranslations
\be
V \ra F(V,\theta,\phi) = V + T(\theta,\phi)
\ee
can be characterised as precisely those transformations that lead to a shell with zero 
pressure and currents. As we will see below, generically they will also generate
accompanying impulsive gravitational waves travelling along the shell. 

The energy density for such a shell is 
\be
\label{bmsshell}
\mu = -\frac1{32m^2\pi}\left(\Delta^{(2)}T - T\right)\;\;,
\ee
and correspondingly the conservation law 
\eqref{Kconslaw} reduces to the evidently correct statement
\be
\del_V\mu = 0\;\;.
\ee
Thus we can define a conserved shell energy $E[T]$ by integrating $\mu$ over
the spatial cross-section of the horizon. In this case, the 1st term
in \eqref{bmsshell} does not contribute and we are left with 
\be
\label{ET}
E[T] = \frac{1}{32\pi m^2}\int\sqrt{|g_{AB}|}d^2 x \;T(\theta,\phi) =\frac{1}{8\pi}\int_{S^2} T(\theta,\phi)
\;\;.
\ee
See the introduction for some discussions of this result.

\item No Matter on the Shell: $S^{ab}=0$

With $F =aV +B$ one finds that the remaining angular components $\gamma_{AB}$ of 
$\gamma_{ab}$ are
\be
\gamma_{AB} = (2/a)(\nabla^{(2)}_A\del_B B - \trac{1}{2}\sigma_{AB}B)\;\;.
\ee
Thus, the requirement that also the energy density $\mu\sim \sigma^{AB}\gamma_{AB}$ vanishes
is
\be
\mu = 0 \quad\Ra\quad \Delta^{(2)} B = B \;\;.
\ee
As there are no (non-singular) solutions to this equation (the eigenvalues of the Laplacian 
are $-\ell(\ell+1) \leq 0$), we conclude that 
\be
\mu=0 \quad\Ra\quad B = 0 \;\;,
\ee
so that the only transformations leading to a shell with vanishing energy-momentum tensor
are the constant rescalings $F(V)=aV$, 
corresponding to the scaling isometry in Kruskal coordinates already mentioned above.

\item Nonexistence of Pure Impulsive Gravitational Waves

With $B=0$, not only the trace of $\gamma_{AB}$ is zero, but evidently $\gamma_{AB}=0$. 
Therefore in this Kruskal horizon shell example we have
\be
S^{ab}=0 \quad\Ra\quad \gamma_{ab}=0\;\;.
\ee
In particular, since the impulsive gravitational wave contributions
are encoded in the transverse traceless part $\hat{\gamma}_{ab}$ of
$\gamma_{ab}$, this means that in the case at hand there can be no pure
impulsive gravitational waves without matter on the shell.

This should be contrasted with the case of null hypersurfaces in Minkowski
space \cite{PenroseSynge} (cf.\ also section 2.4 of \cite{BH}) which, in
the present setting, we can think of as Rindler horizons. In this case
matter and gravitational waves decouple and can exist independently of
each other.

Relaxing the 
assumption that $B$ is smooth, 
one can find (almost) purely gravitational impulsive wave configurations from 
the equation
\be
\mu \sim \Delta^{(2)} B - B = \lambda\delta^{(2)}(\Omega-\Omega_0)
\ee
where the $\delta$-function represents a 
massless point source at fixed angle travelling along ${\mathcal N}$. This case has
been considered in the context of black hole horizons and corrections
to Hawking radiation in \cite{DT1} (cf.\ also \cite{Sfetsos} for generalisations)
and more recently in the context of 
the holography and scattering from black hole horizons e.g.\ in \cite{StSh} and
\cite{Polchinski}. 

It thus appears that the more 
general expression for a shell with matter and impulsive gravitational wave that we
have presented here (e.g.\ the simple configurations arising from BMS transformations)
are smoothed out versions of this singular configuration and can perhaps be employed
in the above contexts.

\item Matter without Impulsive Gravitational Waves: $\hat{\gamma}_{ab}=0$

Finally we determine those soldering transformations that give rise 
to a shell that contains matter but no impulsive gravitational 
waves. We see immediately from \eqref{gammahatK} that a shell with 
no impulsive gravitational wave component must satisfy 
the hyperbolic equation
\be
\nabla^{(2)}_\theta\partial_\theta F 
-\frac1{\sin^2\theta}\nabla^{(2)}_\phi\partial_\phi F = 0
\ee
and the constraint
\be
\nabla^{(2)}_\theta\partial_\phi F = F_{\theta\phi} -\frac{\cos\theta}{\sin\theta}F_\phi = 0\;\;.
\ee
The general solution to these two equations is
\be
\label{Fdip}
F(V,\theta,\phi) = A(V) + B(V)\;\vec{d}.\vec{r}(\theta,\phi)
\ee
for an arbitrary constant vector $\vec{d}$, with $\vec{r}$ the 
unit vector on the 2-sphere,
\be
\vec{r}(\theta,\phi) = (\sin\theta \cos\phi,\sin\theta \sin\phi,\cos\theta)\;\;.
\ee
Thus absence of an impulsive gravitational wave implies that the
coordinate transformation $V_+ = F(V,\theta,\phi)$ only includes monopole
and dipole terms, a plausible result (but we should note that this
simple relation between dipole soldering transformations and 
absence of gravitational radiation holds only in Kruskal coordinates, 
not e.g.\ in Eddington-Finkelstein coordinates).

In particular, choosing $T(\theta,\phi)$ in \eqref{bmsshell} to be an
eigenfunction $Y_{\ell,m}$ of $\Delta^{(2)}$ with $\ell \geq 2$, say,
one obtains a configuration with $p=j_A=0$, but $\mu \neq 0$, accompanied
by an impulsive gravitational wave travelling along the shell.

\end{itemize}

While we only presented this one special case of horizon shells in
detail, it is of course straightforward to apply the procedure outlined
above to other horizon shells, e.g.\ Reissner-Nordstr\o m, or joining
the Schwarzschild black hole to an AdS Schwarzschild black hole etc. In
the cases we have looked at, including extremal Reissner-Nordstr\o m,
we have found no special new features beyond
those already encountered in the above Schwarzschild - Kruskal example.

\section{Horizon Shell in Eddington-Finkelstein coordinates}
\label{secef}

\subsection{General Construction}

We will now briefly also look at the Schwarzschild Horizon Shell in 
Eddington-Finkelstein coordinates, 
\be
ds^2 = -f_{ss}(r)dv^2 + 2 dv dr + r^2 d\Omega^2\quad,\quad f_{ss}(r) = 1-\frac{2m}{r}\;\;.
\ee
In this case it is natural to make the choices $\Phi = r-2m$ and $n=\del_v$. 
Of course one can easily recalculate the $\gamma_{ab}$ from scratch with 
these choices. It is obviously more efficient, however, to make use of the 
results of the previous section and to simply transform them
from Kruskal to Eddington-Finkelstein 
coordinates. However, there is one subtlety that one needs to pay attention
to.

Namely, since Kruskal $V$ and the Eddington-Finkelstein advanced coordinate
$v$ are non-trivially related by 
\be
V= \ex{v/4m}\quad,\quad \frac{\del v}{\del V} = \frac{4m}{V} \;\;,
\ee
so are the corresponding normal vectors, 
\be
\del_V = \frac{\del v}{\del V} \del_v = \frac{4m}{V}\del_v \;\;.
\ee
This leads to an opposite rescaling of the transverse vector $N$, thus of the $\gamma_{ab}$, 
and therefore to
appropriate different 
rescaling of $\mu,j_A,$ and $p$ which are homogeneous in $n$ of degree 0,1,2 respectively
(cf.\ also section 3.11.5 of \cite{poisson2}).
Thus for example, the correct expression for the pressure in Eddington-Finkelstein
coordinates is obtained not just by writing the Kruskal soldering transformation as
\be
V_+= F(V,\theta,\phi) = \ex{v_+/4m} = \ex{f(v,\theta,\phi)}
\ee
and substituting this into \eqref{Kmujp}, but it requires an additional
factor of $V/4m$. Proceeding either way one finds for the $\gamma_{ab}$
\be
\label{EFgamma}
\begin{aligned}
\gamma_{vv} &= \frac{8m}{f_v}\left(\ex{-f/4m}\;\partial_v^2\;\ex{f/4m}\right) - \frac1{2m}\\
\gamma_{vA} &= \frac{8m}{f_v}\left(\ex{-f/4m}\;\partial_v\partial_A\;\ex{f/4m}\right)\\
\gamma_{AB} &= \frac{8m}{f_v}\left(\ex{-f/4m}\;\nabla^{(2)}_A\partial_B\;\ex{f/4m}
 +\frac12\sigma_{AB}(f_v -1)\right)
\end{aligned}
\ee
and then 
\be
\begin{aligned}
\label{EFmujp}
\mu &= -\frac1{8m\pi f_v}\left(\ex{-f/4m}\Delta^{(2)}\ex{f/4m} + f_v - 1\right)\\
j^A &= \frac1{64m^2\pi}\sigma^{AB}\left(2\frac{f_{vB}}{f_v} + \frac{f_B}{2m}\right)\\
p &= -\frac1{8\pi}\left(\frac{f_{vv}}{f_v} + \frac{f_v}{4m} - \frac1{4m}\right)\;\;.
\end{aligned}
\ee

The one non-trivial conservation law \eqref{conslaw} is now
\be
\label{efcons}
(\partial_v + \tilde{\kappa})\mu
+ (\nabla^{(2)}_A +\frac12\gamma_{Av})j^A +\frac14(g^{AB}\gamma_{AB}-\frac4m)p = 0\;\;,
\ee
which can be obtained directly from \eqref{conslaw} or alternatively by transforming
\eqref{Kconslaw} according to the above prescription. 
This equation is again of course identically satisfied though the calculation to check 
this is somewhat lengthier than that in Kruskal coordinates. The angular components
of the conservation law are just the obvious Eddington-Finkelstein counterpart 
of \eqref{Kangle}.

\subsection{Special Cases}

\begin{itemize}
\item BMS Supertranslations of $v$

One potentially interesting special class of soldering transformations 
are horizon BMS supertranslations of $v$ \cite{andy2, Hawking}
\be
v \ra f(v,\theta,\phi) = v + t(\theta,\phi)
\ee
In this case 
the energy density, current and pressure are
\be
p=0\quad,\quad j^A = \frac1{128m^3\pi}\sigma^{AB}\partial_B t\quad,\quad
\mu = -\frac1{8m\pi}\ex{-t/4m}\Delta^{(2)}\ex{t/4m}
\ee
Even though the conservation law \eqref{efcons} is non-trivial in this case, 
because of the presence of both $\mu$ and the currents $j^A$, it is
nevertheless true (and evident from the above expression for $\mu$), that
the energy density is conserved along the shell, $\del_v\mu=0$, as in the 
case of Kruskal horizon supertranslations discussed in section \ref{subk}.
Therefore we can again define an associated conserved energy $E[t]$ given
(after an integration by parts) by
\be
\label{efet}
E[t] = \int \sqrt{|g_{AB}|}d^2x\;\mu = -\frac{1}{8\pi}\int_{S^2}\frac{1}{4m}\sigma^{AB}\del_At \del_Bt \;\;.
\ee 
Since, when translated back to Kruskal coordinates, this transformation
\be
V_+ = V\ex{t/4m},
\ee
is not of the dipole form \eqref{Fdip}, this energy density and flux will almost 
invariably be accompanied by an impulsive gravitational shock wave.

\item Comparison to calculations in \cite{BI}

As our final illustration of the formalism, 
we look at a special case of the general formulation 
to make contact with some simple solderings considered in \cite{BI}.
The general setting is that of two spherically symmetric space-times 
\be
\dd s_\pm^2 = -\ex{2\psi_\pm}f_\pm\dd v^2 + 2\ex{\psi_\pm}\dd v\dd r + 
r^2\dd\Omega^2.
\ee
joined along their (common) static horizon at $r=r_s$, where
$f_\pm(r_s)=0$. Restricting to the case for which the interior and
exterior geometries are those of the Schwarzschild black hole, $f_\pm(r)
= f_{ss}(r)$, and choosing the exterior metric to be of the standard form
(with $\psi_-=0$), the non-trivial soldering is encoded in a non-trivial
$\psi_+(v)=\psi(v)$, leading to a discontinuity of the metric in these coordinates.
In \cite{BI} it is then shown how the energy-momentum tensor
of the shell can be determined from the transverse extrinsic curvatures of the
metrics on the 2 sides. 

Alternatively, this interior form of the metric can be obtained by taking the interior metric
in standard Eddington-Finkelstein coordinates ($\psi=0$) and then applying the soldering transformation
\be
v_+ = f(v) = \int^v \ex{\psi}\dd v\;\;.
\ee
From \eqref{EFmujp} one then deduces immediately that this shell carries 
non-zero energy density and pressure given by
\be
\mu = \frac1{8\pi m}[\ex{-\psi}],\quad p = -\frac1{8\pi}[\partial_v\psi + \kappa_0\ex{\psi}]
\ee
($\kappa_{0}=1/4m$ is the surface gravity of the Schwarzschild black hole), 
in complete agreement with equation (68) of \cite{BI}.
\end{itemize}

\subsection*{Acknowledgements}

We are grateful to Andy Strominger for his comments on a preliminary
version of this article. MO acknowledges financial support from the
EU-COST Action MP1210 ``The String Theory Universe'' and from the
Albert Einstein Center for Fundamental Physics, Bern. MO would like to
thank the organisers and participants of the ICTP Conference on ``Recent
Progress in Quantum Field Theory and String Theory'' in Yerevan, Armenia
(September 14th - 19th 2015) where a very preliminary version of this
work was presented.
The work of MB is partially supported through the NCCR SwissMAP (The Mathematics
of Physics) of the Swiss Science Foundation.
The research of MO was funded by the Slovenian Research Agency.

\rnc{\Large}{\normalsize}


\begin{thebibliography}{00}
\addcontentsline{toc}{section}{References}
\frenchspacing
\begin{small}
\addtolength{\itemsep}{-4pt}

\bibitem{BI} C.\ Barrabes, W. Israel, \textit{Thin shells in general relativity and 
cosmology: The lightlike limit}, Phys.\ Rev.\ D43 (1991) 1129-1142.

\bibitem{BH}
C.\ Barrabes, P.\ Hogan, \textit{Singular Null Hypersurfaces in General Relativity}
(World Scientific, 2003).

\bibitem{poisson1} E.\ Poisson, \textit{A reformulation of the Barrabes-Israel null-shell formalism}, 
\texttt{arXiv:gr-qc/0207101}.

\bibitem{poisson2} E.\ Poisson, \textit{A Relativist's Toolkit} (Cambridge University Press, 2004).

\bibitem{DT2} T.\ Dray, G.\ 't Hooft, \textit{The Effect of Spherical Shells of Matter and
the Schwarzschild Black Hole}, Commun.\ Math.\ Phys.\ 99 (1985) 613-625.

\bibitem{dg1} C.\ Duval, G.\ Gibbons, P.\ Horvathy, \textit{Conformal Carroll groups and BMS symmetry}, 
\texttt{arXiv:1402.5894 [gr-qc]}.

\bibitem{dg2} C.\ Duval, G.\ Gibbons, P.\ Horvathy, \textit{Conformal Carroll groups}, 
\texttt{arXiv:1403.4213 [hep-th]}.

\bibitem{ih1} A.\ Ashtekar, B.\ Krishnan,
\textit{Isolated and dynamical horizons and their applications},
Living Rev. Relativity 7,  (2004) 10. \texttt{http://www.livingreviews.org/lrr-2004-10},
\texttt{arXiv:gr-qc/0407042}. 

\bibitem{ih2} 
J.\ Engle, T.\ Liko, \textit{Isolated horizons in classical and quantum gravity},
\texttt{arXiv:1112.4412 [gr-qc]}.

\bibitem{koga}
J.\ Koga, \textit{Asymptotic symmetries on Killing horizons}, Phys.\ Rev.\ D64,
(2001), 124012, \texttt{arXiv:gr-qc/0107096}.

\bibitem{DT1}
T.\ Dray, G.\ 't Hooft, \textit{The gravitational shock wave of a massless
particle}, Nucl.\ Phys.\ B253 (1985), 173-188.

\bibitem{StSh}
S.\ Shenker, D.\ Stanford, \textit{Black holes and the butterfly effect}, 
JHEP 1403 (2014) 067, \texttt{arXiv:1306.0622 [hep-th]}.

\bibitem{Polchinski} J.\ Polchinski, \textit{Chaos in the black hole S-matrix}, 
\texttt{arXiv:1505.08108 [hep-th]}.

\bibitem{andy2}  A.\ Strominger, A.\ Zhiboedov, \textit{Gravitational Memory, BMS Supertranslations and
Soft Theorems}, \texttt{arXiv:1411.5745 [hep-th]}.

\bibitem{Hawking} S.\ Hawking, \textit{The Information Paradox for Black Holes}, 
\texttt{arXiv:1509.01147 [hep-th]}; S.\ Hawking, M.\ Perry, A.\ Strominger
(to appear)


\bibitem{wz} R.\ Wald, A.\ Zoupas, \textit{A General Definition of "Conserved Quantities" in General Relativity
and Other Theories of Gravity}, Phys.Rev. D61 (2000) 084027, \texttt{arXiv:gr-qc/9911095}.

\bibitem{bt1} G.\ Barnich, C.\ Troessaert, \textit{BMS charge algebra}, 
JHEP12(2011)105, \texttt{arXiv:1106.0213 [hep-th]}.

\bibitem{andy1} A.\ Strominger, \textit{On BMS Invariance of Gravitational Scattering}, 
JHEP07(2014)152. \texttt{arXiv:1312.2229 [hep-th]}.

\bibitem{dggp} L.\ Donnay, G.\ Giribet, H.\ Gonzalez, M.\ Pino, 
\textit{Super-translations and super-rotations at the horizon}, 
\texttt{arXiv:1511.08687 [hep-th]}.

\bibitem{NP} Y.\ Nutku, R.\ Penrose, \textit{On Impulsive Gravitational Waves}, Twistor Newsletter 34
(1992) 9-12 (available from \texttt{http://people.maths.ox.ac.uk/lmason/Tn/}).


\bibitem{xbkm} X.\ Bekaert, K.\ Morand, \textit{Connections and dynamical trajectories in generalised
Newton-Cartan gravity II. An ambient perspective}, \texttt{arXiv:1505.03739 [hep-th]}.

\bibitem{jelle} J.\ Hartong, \textit{Gauging the Carroll Algebra and Ultra-Relativistic Gravity}, 
\texttt{arXiv:1505.05011 [hep-th]}.

\bibitem{PenroseSynge}
R.\ Penrose, \textit{The Geometry of Impulsive Gravitational Waves}, in 
\textit{General Relativity, Papers in Honour of J.L. Synge},\ (Clarendon
Press, Oxford, 1972), p.101. 

\bibitem{Sfetsos} K.\ Sfetsos, \textit{On Gravitational Shock Waves in Curved Spacetimes},
Nucl.Phys. B436 (1995) 721-746, \texttt{arXiv:hep-th/9408169}.

\end{small}
\end{thebibliography}
\end{document}